# Calipso: Physics-based Image and Video Editing through CAD Model Proxies

Nazim Haouchine[1], Frederick Roy[1], Hadrien Courtecuisse[2], Matthias Nießner[*3], and Stephane Cotin[†1]

[1]Inria, Mimesis Group
[2]CNRS, University of Strasbourg
[3]Technical University of Munich

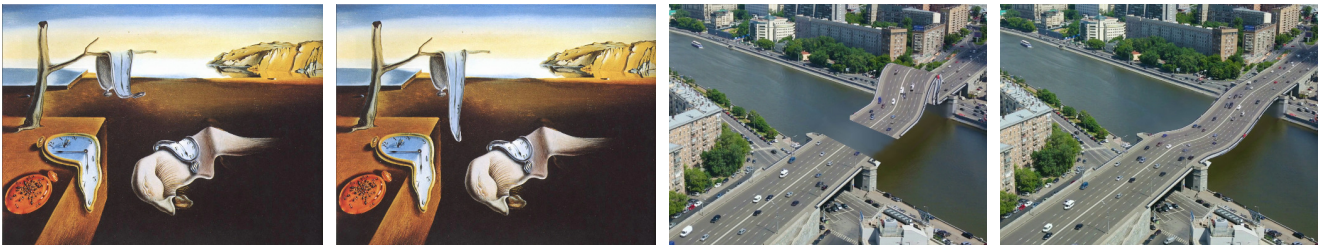

**Figure 1:** *Our method allows a user to interact in a 3D manner with objects in images and videos by producing rigid and deformable transforms, topological changes and physical attribute editing (e.g., mass, stiffness, gravity). Using an intuitive mesh refinement and 3D/2D alignment approach and estimating dynamics from image flow, our system produces a final composition without cumbersome user input, while preserving visual consistency. Left: editing Dali's painting by interactively pulling the watch or decreasing its stiffness leading to melting it. Right: bending a car bridge by pulling it from a side or by progressively increasing its mass.*

## Abstract

We present Calipso, an interactive method for editing images and videos in a physically-coherent manner. Our main idea is to realize physics-based manipulations by running a full physics simulation on proxy geometries given by non-rigidly aligned CAD models. Running these simulations allows us to apply new, unseen forces to move or deform selected objects, change physical parameters such as mass or elasticity, or even add entire new objects that interact with the rest of the underlying scene. In Calipso, the user makes edits directly in 3D; these edits are processed by the simulation and then transferred to the target 2D content using shape-to-image correspondences in a photo-realistic rendering process. To align the CAD models, we introduce an efficient CAD-to-image alignment procedure that jointly minimizes for rigid and non-rigid alignment while preserving the high-level structure of the input shape. Moreover, the user can choose to exploit image flow to estimate scene motion, producing coherent physical behavior with ambient dynamics. We demonstrate Calipso's physics-based editing on a wide range of examples producing myriad physical behavior while preserving geometric and visual consistency.

## 1 Introduction

Image and video editing is a core research problem with a long history in computer graphics and vision, impacting a wide range of applications such as movie production post-processing, life-style magazine content polishing, and many more. Along with impressive research advances, commercial tools (e.g., Adobe Photoshop) have been developed that put theory into practice, and have made many sophisticated editing algorithms popular and available to the masses. Early edit operations were fairly low-level, and involved significant effort to achieve desired results while relying on skilled users in order to ensure seamless edits. The key towards an easier editing pipeline is finding the right abstraction, allowing edit manipulations to become more high level tasks. Seam Carving [Avidan and Shamir 2007] for content-aware resizing or PatchMatch [Barnes et al. 2009] for content-aware filling are just a few of many great achievements in this context.

Despite significant progress towards making 2D content manipulation easier and accessible, it is still challenging for novice users to achieve high-quality editing results in many situations. One of the main reasons is that images and videos are only given as a 2D projection of a 3D environment, where the information of the underlying 3D scenes – including geometry, material, and lighting – is not fully reflected. It becomes even more challenging to apply edits in a *physically-consistent* manner. For instance, adding new objects that interact with the rest of the scene, or simply increasing the mass or changing the elasticity of an object, is inherently difficult to achieve in a purely image-based domain.

In this work, we address these challenges by introducing Calipso, a real-time editing method focused specifically on physics-based modifications of images and videos. The core idea is to first abstract the target content with a 3D representation. This abstraction serves the user as an interaction proxy, and is used to enforce physics consistency. Thus, edits can be directly applied in 3D, and we can simulate a desired physics behavior. Calipso then transfers the outcome to the underlying 2D content through shape-to-image correspondences in a photo-realistic rendering process to generate the modified result.

In order to obtain the 3D geometry for an object of interest, we introduce a novel model fitting interface that allows the selection and alignment of shapes from a model database. The user only needs to place a CAD model roughly on top of the target area and provide a sketch of the silhouette of the object of interest. Based on this input, we formulate an energy minimization problem that jointly aligns and deforms the CAD models such that the target projections match while preserving the global 3D structure of the input shape, even if the CAD model's original shape is changed significantly.

---

[*]niessner@tum.de
[†]stephane.cotin@inria.fr

With this 3D abstraction, we introduce *physics-based* editing by running a full physics simulation using the obtained geometric proxies. Calipso provides an interface where edits can easily be realized. This new paradigm of physics-based editing not only facilitates non-rigid edits and changes in an object's appearance according to an interactive finite element simulation, but also supports manipulations in physical behavior; for example, mass, velocity, or other physics parameters of an object can easily be adjusted to achieve different visual effects in the output video.

Calipso makes it easy to realize these edits with *interactive* feedback with little manual intervention, and is based on two main contributions:

- We realize physics-based image and video editing through CAD model proxies supported by a finite element simulation, allowing for a wide range of manipulations, including rigid transforms, geometry and topological changes, elastic deformations, collision, and physical parameter updates such as mass, stiffness, damping, or gravity.

- We introduce an efficient CAD-to-image alignment procedure that jointly minimizes for rigid and non-rigid model alignment while preserving the high-level structure of the input shape.

## 2 Related Works

**Traditional Image Editing** Image editing methods have a long history in the research community; here, we will review the most relevant works to our approach. Much of this research is inspired by commercial tools. A prime example is Adobe Photoshop, which allows a user to perform a large variety of both low- and high-level image editing operations in an interactive UI. Many state-of-the-art image editing approaches have been integrated into Photoshop, such as content-aware filling using the PatchMatch inpainting method [Barnes et al. 2009] or Seam Carving for content-aware resizing [Avidan and Shamir 2007]. However, there are many other powerful approaches to facilitate image manipulation. One direction is to incorporate semantics, and perform operations on 2D objects rather than on the per-pixel level in order to support scaling, stretching, bending, etc. [Barrett and Cheney 2002]. Another direction is editing the appearance of objects in an image. For instance, Fang et al. [2004] combine texture synthesis and shape-from-shading to modify the textures of existing objects, and Khan et al. [2006] alter the material properties of objects, adding transparency, translucency and gloss based on insights from human perception.

Oh et al. [2001] enable image-based modeling and photo editing by assigning depth layers to perform viewpoint changes in a photograph. This idea is further extended by Chen et al. [2011] to work on videos. Another possibility is to incorporate content from images from the web, allowing insertion of new objects into selected regions [Goldberg et al. 2012]. Chen et al. [2009] propose Sketch2Photo, which composes multiple photos based on user-provided sketches in a seamless manner.

These methods achieve impressive results, and have received wide attention; however, their fundamental limitation is the inherent lack of understanding of underlying 3D geometry, which makes the incorporation of physics-based edits challenging.

**Image Editing with Domain Knowledge** Editing images can be supported by leveraging domain knowledge in order to constrain a parameter space that force results to stay within realistic bounds. For instance, Debevec et al. [1996] model and render architectures from photographs. They use regular symmetric structures of architectural models to reveal novel views of buildings. Another important domain is human faces: Blanz and Vetter [1999] build a low-dimensional parametric face model that allows the synthesis of 3D faces. Face2Face [Thies et al. 2015; Thies et al. 2016] extends this by allowing photorealistic edits of facial expressions in videos in real time. Human shapes is another domain, addressed by MovieReshape [Jain et al. 2010] which enables quick and easy manipulation of human bodies in 2D content based on an underlying 3D morphable model. Bai et al. [2012] focus on temporal edits by selectively de-animating videos. Their core idea is to compute a warp field constraint by a set of manually-provided strokes that removes the large-scale motion in these regions while leaving finer-scale, relative motions intact.

**Object-centric Image Manipulation** The most related line of work to ours is the object-centric manipulation of images. Photo Clip Art [Lalonde et al. 2007] inserts objects from a clip art database whose shape have all required properties, including camera pose, lighting, resultion, etc.. The placement of the inserted objects is 3D; however, the source objects are from the LabelMe dataset [Russell et al. 2008] where ground truth object segmentations are provided. Objects can also be inserted into an existing photograph by rendering a synthetic CAD model on top of the image with subsequent compositing [Debevec 2008]. In order to facilitate compositing and generate realistic results, the scene's illumination must be captured. For this pupurose, Debevec et al. [2008] utilize a mirrored sphere. A similar idea can be applied in order to insert synthetic objects into legacy photographs. Karsch et al. [2011] first annotate the geometry and lighting, and then render 3D objects into the scene and run a final compositing step. These methods produce impressive results; however, they focus on inserting new content into images rather than modifying the appearance of existing objects.

Zheng et al. [2012] addresses the task of editing objects in a scene by cuboid proxies. First, they compute an image composition and abstract scene objects with models of 3D cubes. Their edits focus on the scene level and allow for a variety of editing tasks; e.g., replace all furniture in a living room with different models and embed them (naturally) in the image. Chen et al. [2013] follow up on this direction with an impressive editing framework. They propose 3-sweep, an interactive editing tool, which facilitates the extrusion of coarse 3D shapes from images with only three sweep strokes, each defining a dimension. Once a 3D object is extracted, it can be quickly edited and placed into photos of 3D scans. This facilitates a large variety of powerful editing operations; however, in contrast to our work, 3-sweep's focus is on rigid transformations of extracted objects and associated parts. Instead of extruding the 3D geometry directly from the image, Kholgade et al. [2014] utilize 3D CAD models from a shape database as a prior. They manually retrieve and align 3D models to an image, and estimate illumination and texture of the aligned model. The CAD model then serves as an interaction proxy to facilitate image edits by transferring modifications of the 3D shape proxy to the image. While they mostly focus on image examples with rigid object edits, their method also allows for non-rigid object deformation with an as-rigid-as-possible editing constraint [Sorkine and Alexa 2007]. Our method uses a similar pipeline; however, our goal is to enable physics-based editing of image and video content by connecting 3D CAD model proxies with a faithful physics simulation. This not only allows us to perform interactive edits constrained by a true physics simulation, but also modify the simulation results and image/video appearance by changing the underlying physics parameters; e.g., mass, mate-rial, elasticity, etc.. In addition, we propose a non-rigid alignment framework based on silhouette constraints compared to the alignment tool by Kholgade et al. that requires heavy user interaction to specify all constraints in a tedious manual procedure. Bazin et al. [2016] also argue for a physics simulation to edit videos. Their main idea for the physics simulation is to discourage object parts with no texture information from becoming visible. In contrast to our method, their 3D proxies are essentially modeled around the input video, limiting the complexity of their results. Davis et al. [2015] extract an image-space representation of object structure from video, and use it synthesize physically plausible animations. Their approach facilitates edits based on unseen physical forces without knowledge of the scene geometry or material properties.

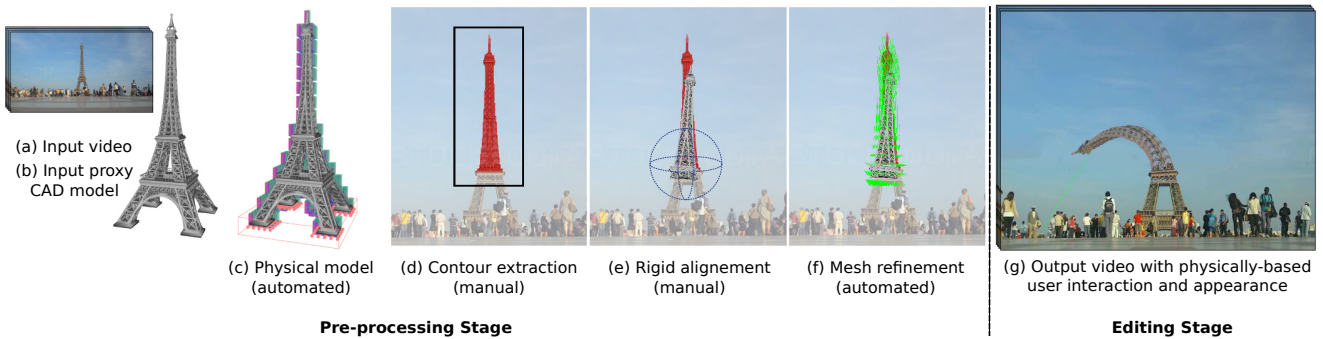

**Figure 2:** *Problem setup and overview: an input video Q (a) and a proxy CAD model S (b) are expanded using a deformable model built on the fly (c). The user manually defines the object contour through an intuitive graph-based foreground extraction (d), and roughly aligns the CAD model on the image (e) to obtain the camera position $\mathbf{P}$. Our method automatically refines the mesh and aligns it non-rigidly with the image by computing $\mathscr{T}(S)$ through an energy minimization process (f). Finally, the user can interact with the targeted object (g) by manipulating its geometry and by editing its local and global physical properties gathered in the set $\phi$ and the function $\mathbf{F}$. The resulting composite video $Q^*$ maintains visual fidelity $\mathscr{A}$ using image-based inverse rendering.*

## 3 Problem Setup and Overview

Our system aims at enabling physics-based user-edits on videos or images using a 3D CAD model proxy (available in various internet databases). As shown in the pipeline of Figure 2, our system is separated onto two stages, a pre-processing stage and an editing stage. In the pre-processing stage a physical model in automatically built upon the input CAD model. This physical model is used for both the CAD-to-model alignement and the editing stage. The user then defines the 2D object's contour, in order to be used to automatically align the model in the image while simultaneously refining the 3D mesh geometry. Once aligned, and object's mesh is refined, the appearance is estimated to enable consistent rendering. The user can now edit the video in 3D using various types of manipulations, while relying on the object's intrinsic and extrinsic physical characteristics and scene dynamics and maintaining a geometrically and visually consistent scene.

Formally speaking, let $Q = \{I_t, I_{t+1} \cdots, I_{t+k}\}$ be an image sequence, where $I$ is an image of size $w \times h$ and $t$ the time. And let $S$ be a 3D non-textured surface mesh composed of vertices, faces and normals that semantically correspond to an object in the sequence $Q$. We seek to produce a new image sequence $Q^* = \{I_t^*, I_{t+1}^* \cdots, I_{t+k}^*\}$, which enables the user to manipulate desired objects by editing their geometrical and physical behavior. The sequence $Q^*$ can be determined through the editing function $\Gamma$:

$$Q^* = \Gamma(Q, S, \mathbf{P}) \qquad (1)$$

where $\mathbf{P}$ is the camera projection matrix relating the 3D model in world coordinates to its 2D projection in pixel coordinates. The camera is defined by the user during the initial alignment and assumed to be fixed during the sequence.

In practice, the 3D model geometry $S$ obtained through proxy does not fit the targeted object in the sequence. Thus, we define a transform function $\mathscr{T}$ that automatically updates the 3D model shape using its projected contour on the image. The contour is user-defined and enables inpainting the sequence $Q_p$, revealing new parts. We obtain a new model $\mathscr{T}(S) = S^*$ that faithfully matches the object in the image.

In addition to the transform function, we consider an appearance function $\mathscr{A}$ that estimates object appearance in terms of texture and illumination to produce a seamless final composite sequence $Q^*$. This amounts to estimating illumination and diffuse reflectance by minimizing a similarity function between the original image and the rendered one such that $\mathscr{A}(S^*, Q_p) = Q$.

Finally, we add a set of physical parameters which govern the scene dynamics, such as gravity, friction or external pressure. These quantities are gathered by the function $\mathscr{F}$, and can be set and edited by the user, as well as estimated from the video. In fact, when editing videos, it can be very difficult and time consuming to maintain visual consistency while simultaneously change an object's behavior. By exploiting video dynamics to extract velocity, acceleration, and moment of collisions, we can significantly improve this consistency. We define the set $\phi$ as the set of the object physical properties (e.g., mass, damping, stiffness), which are set by the user and can be modified interactively.

Overall, we can rewrite the editing function $\Gamma$ as

$$Q^* = \Gamma(\mathscr{A}(S^*, Q_p), S_{\{\mathscr{F}, \phi\}}, \mathbf{P}) \qquad (2)$$

Our method involves determining $Q^*$ efficiently without cumbersome user input to allow interactive physics-based editing of an input video sequence $Q$. We exploit the physical model introduced in Section 4 to correct the initial geometry while satisfying both physical and projective camera constraints. This permits simultaneously computing the new model $S^*$ and alignment on the image. This process is described in Section 5. Once $S^*$ is computed and aligned, the appearance $\mathscr{A}$ is estimated through a similarity-based optimization process (see Section 6). The user can finally interact with the sequence by editing its physics, geometry, and position in space and time, while relying on physical cues extracted from the videos, such as motion velocity and instant of collision (describe in Section 7).

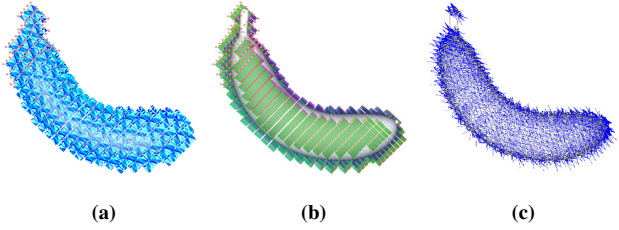

**Figure 3:** *FEM volume, boundary conditions, and mapping: the volumetric mesh V is built upon the surface model S following several representations: (a) a grid tetrahedron volume model, (b) an grid hexahedron volume model, and (c) the volume and the surface mapping. The red spheres represent the boundary constraints.*

# 4 Physics-based Model

The literature related to physical models is vast and crosses many scientific domains such as engineering, computational mechanics and computer graphics [Nealen et al. 2006]. While dedicated models such as thin-plate splines [Bookstein 1989], free-from [Coquillart 1990], or as-rigid-as-possible [Sorkine and Alexa 2007] have proved their relevance and efficiency for user-guided model-to-image fitting and isometric deformations, we intend to have a more generic model. In our method, physical models are used in three ways:

1. to facilitate the model-to-image alignment, where the input 3D surface does not correspond to the object in the sequence and mandates deforming its geometry,

2. to interactively manipulate objects in a 3D manner including stretching, torsion, compression.

3. to update physical attributes whether intrinsic (related to the object, like stiffness, mass, and damping), or extrinsic (related to the scene, like gravity, motion dynamics, and surrounding forces).

In order to satisfy the three above-mentioned needs, we rely on a physically-based model that permits an automatic geometry correction, an interactive user interaction and physical properties update. Several characteristics are sought for our deformable model: a low computational cost while maintaining reasonable accuracy, the ability to handle large deformations, and a low-dimensional parameterization. For this purpose, a Saint-Venant Kirchoff material appears to be a relevant strategy: the model is non-linear, thus allowing handling of geometric non-linearities (torsion, compression, elastic deformation); it can be quickly computed following the work of Kikuuwe *et. al.* [Kikuuwe et al. 2009]; and it relies on few material parameters, Young's modulus $E$ as a measure of the stiffness of the material and Poisson's ratio $\nu$ as an estimate of its compressibility. $E$ and $\nu$ can be tuned to respectively increase or decrease the locality of the deformations and control the volume preservation, while enforcing the consistency of the geometry (even for large deformations).

## 4.1 Discretization with Finite Element Method

Without loss of generality, we use the Finite Element Method (FEM) to discretize partial differential equations of solid continuum mechanics. This discretization is computed on a volumetric mesh $V$ with a finite number of degrees of freedom (element's nodes). This volume representation $V$ is built on-the-fly from a voxelization of the input 3D model $S$ (cf. Figure 3 (a) and (b)). The deformable object is represented as a volumetric mesh composed of a sparse grid of tetrahedral or hexahedral elements. The number of elements are to be chosen adequately in order to ensure interactive performance as well as sufficient accuracy.

A particular object deformation is specified by the displacements of nodal positions and the nodal forces. In general, the relationship between nodal forces and nodal positions is non-linear. When linearized [Courtecuisse et al. 2014], the relationship for an element $e$ connecting $n_e$ nodes can simply be expressed as

$$\mathbf{f_e} = \mathbf{K_e}\delta\mathbf{u_e} \quad (3)$$

where $\mathbf{f_e} \in \mathbb{R}^{3n_e}$ contains the $n_e$ nodal forces and $\delta\mathbf{u_e} \in \mathbb{R}^{3n_e}$ the $n_e$ nodal displacements of an element. The matrix $\mathbf{K_e} \in \mathbb{R}^{3n_e \times 3n_e}$ is called the stiffness matrix of the element. Because forces coming from adjacent elements add up at a node, a stiffness matrix $\mathbf{K} \in \mathbb{R}^{3n \times 3n}$ for an entire mesh with $n$ nodes can be formed by assembling the element's stiffness matrices $\mathbf{K_e}$. The equation of deformation of an object will therefore take the general form:

$$\mathbf{K}\delta\mathbf{u} = \mathbf{f} \quad (4)$$

where $\mathbf{f}$ are external forces and $\delta\mathbf{u}$ represent displacements of nodal positions of the whole volume $V$. The computation of the stiffness matrix $\mathbf{K}$ is nonlinear due to the non-linearity of Green-Lagrange strain tensor and is built depending on the material properties, Young's modulus and Poisson's ratio.

## 4.2 Mapping and Force Propagation

We distinguish in the paper the physical model and the visual model. The volumetric mesh $V$ is composed of $n$ nodes $\mathbf{u} \in \mathbb{R}^{3n}$, representing the mechanical model whereas $S$, the surface mesh is composed of $b$ vertices $\mathbf{s} \in \mathbb{R}^{3b}$ (with generally $n < b$) and represents the visual model. Both models are linked with a bi-directional mapping (cf. Figure 3 (c)):

Let $\mathbf{J}$ be the Jacobian matrix used to map the nodes positions $\mathbf{u}$ of the volume model to the vertices positions $\mathbf{s}$ of the surface model. The positions can be mapped following $\mathbf{x}_u = \mathbf{J}\mathbf{x}_s$ where $\mathbf{J} = \frac{\partial \mathbf{x}_t}{\partial \mathbf{x}_p}$. Using $\mathbf{J}$ the velocities are mapped as: $\mathbf{v}_u = \mathbf{J}\mathbf{v}_s$, where accelerations can be mapped using $\mathbf{a}_t = \mathbf{J}\mathbf{a}_p + \frac{\partial \mathbf{J}}{\partial \mathbf{x}_p}\mathbf{v}_p$. The matrix $\mathbf{J}$ contains barycentric coordinates of the degrees of freedom of surface w.r.t to the corresponding volume, which we assume remains valid during the deformation.

The stress/strain relation related to the physical model implies that each displacement generates forces. The propagation of the positions and velocities are from the surface to the volume. The forces are propagated conversely, from the volume to the surface. Given forces $\mathbf{f}_u$ applied to the volume model, the mapping computes and accumulates the equivalent forces $\mathbf{f}_s$ applied to its surface. Since equivalent forces must have the same power, the following relation holds: $\mathbf{v}_s^\mathsf{T}\mathbf{f}_s = \mathbf{v}_u^\mathsf{T}\mathbf{f}_u$ Given $\mathbf{J}$ and using the virtual work principle, the previous relation can finally be expressed as $\mathbf{f}_s = \mathbf{J}^\mathsf{T}\mathbf{f}_u$

It is important to recall that only a part of the volume deform during the alignment, since the contour are 2D projections; the deformation is thereby driven by loads acting solely on its surface (aside from gravity). The mapping relations permit propagating this deformation to the rest of the model, out-of-plane, where we can ensure a consistent final mesh $S^*$.

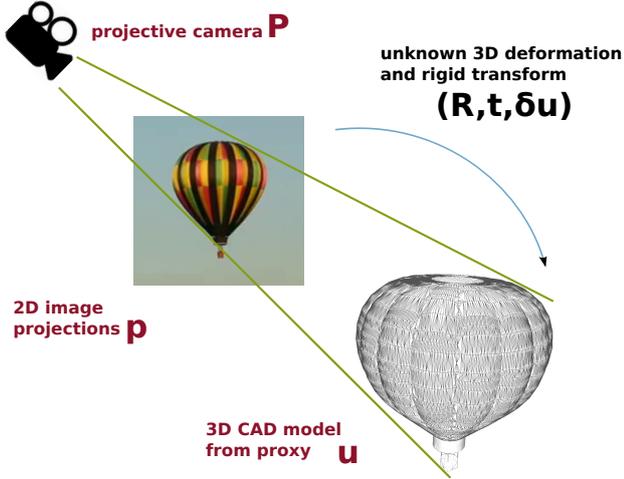

**Figure 4:** *Model-to-image alignment: we seek to find the rotation* **R**, *the translation* **t**, *and the deformation* δ**u** *between the 2D image projections and the 3D model, with respect to the projection camera* **P**. *Finding this transformation will simultaneously result in a refined 3D model $S^*$ and a coherent model-to-image alignment through an automated process.*

## 5 Model-to-image Alignment

In order to align the 3D model $S$ on the image $I$, one must first compute the transformation $\mathscr{T}$, that encodes both rigid and non-rigid transformations, so that $\mathscr{T}(S) = S^*$. Once computed, the new deformed model $S^*$ can be projected on each image composing the sequence following the projection matrix **P**. Since no 3D information about the shape of our targeted objects is provided, we rely only on 2D projections to both correct object geometry and align it on the image. More precisely, if we denote $\mathbf{p} = \{p_i \in \mathbb{R}^2\}, i \in \{1 \ldots m\}$ the positions of these 2D projections, the transformation $\mathscr{T}$ can be written as:

$$\mathbf{P} \cdot \mathscr{T}^{\mathbf{R},\mathbf{t},\delta\mathbf{u}}(\mathbf{u}) = \mathbf{p} \quad (5)$$

where **R** and **t** are respectively the $3 \times 3$ rotation matrix and $3 \times 1$ translation vector that represent the rigid transformation, and δ**u** is the $3n$ vector of nodal displacements which encodes the deformations of the physical model $V$. This formulation is illustrated in Figure 4.

Finding $\mathscr{T}$ is a non-trivial task, since several 3D shapes can lead to the same 2D projection. In a purely rigid scenario, this process is known as a *perspective-n-point* problem [Lepetit et al. 2008] and can be solved efficiently given a set of correspondences between 3D points and their 2D projections. In our case, the problem is more complex since dealing with deformable objects leads to a higher number of degrees of freedom. To this end, we express the problem as an energy minimization problem where the solution is the equilibrium between the internal forces of the physical model **u** and the external forces resulting from the projection of the object on the image **p**, while satisfying the projection constraints of the camera **P**.

**Object contours as physical penalties** We propose to consider the contours of the object as external penalties that will constrain the mechanical model to fit its projection of the image. The contour is extracted using the well-known graph-based foreground extraction algorithm, grabcut [Rother et al. 2004]. By allowing the user to define the inner and outer parts of the object, a segmentation is computed. We obtain from this segmentation the $2m$ vector **p** that represents the contour. This contour is smoothed over a Gaussian function, following a moving least-square process [Levin 2004] to removed scattered and possible noise generated from the segmentation. This permits avoiding local minima during the deformable alignment. The contour is finally integrated in the physical system as a stiff force acting on the deformable volume to enforce the alignment by fitting the 2D contours non-rigidly, following

$$\sum_i^m \frac{1}{2}\kappa\|p_i - \mathbf{R}u_i - \mathbf{t}\|^2 \quad (6)$$

where $\kappa$ is the stiffness coefficient of the contour shape, and is typically of the same orders of magnitude as the Young's Modulus.

**Volume forces as internal energy** We consider in the global minimization the model's volume as internal energy to ensure consistent and realistic deformations. This energy is computed from the general formulation of Eq. 4 describing object's stiffness as follows:

$$\frac{1}{2}\|\delta\mathbf{u}^\top \mathbf{K}\delta\mathbf{u}\| \quad (7)$$

**Projective constraints** We include in the system a projective constraint that corresponds to the camera position. This constraint ensures that resulting points **u** lie on sightlines passing through camera position (cf. Figure 5). Mathematically, this amounts to minimizing the following reprojection error:

$$\|\frac{\mathbf{P}_1 u_i}{\mathbf{P}_3 u_i} - x_i; \frac{\mathbf{P}_2 u_i}{\mathbf{P}_3 u_i} - y_i\| \quad \text{for } i = 1, \ldots, m. \quad (8)$$

where **P** is the projection matrix and $\mathbf{P}_k$ is its $k^{th}$ row and $x_i$ and $y_i$ the pixel coordinates of each $p_i$.

**Energy minimization** Finally, the cost function is written such that: 1) it satisfies the boundary constraints, where the projection of the mesh on the images should result in a minimal distance error, 2) it satisfies the camera pose constraints, and 3) it reaches equilibrium between the internal and external forces of the physical model. This leads to the following expression

$$\min_{\mathbf{R},\mathbf{t},\delta\mathbf{u}} \left( \overbrace{\frac{1}{2}\|\delta\mathbf{u}^\top \mathbf{K}\delta\mathbf{u}\|}^{\text{Physical energy}} + \overbrace{\sum_{i \in m} \frac{1}{2}\kappa\|p_i - \mathbf{R}u_i - \mathbf{t}\|^2}^{\text{Contours fitting}} \right)$$
$$\text{s.t. } \underbrace{\|\frac{\mathbf{P}_1 u_i}{\mathbf{P}_3 u_i} - x_i; \frac{\mathbf{P}_2 u_i}{\mathbf{P}_3 u_i} - y_i\| \quad \text{for } i = 1, \ldots, m.}_{\text{Projective constraints}} \quad (9)$$

The question of initialization naturally arises from Eq. 9, where the convergence of the system is sensitive to the camera pose **P**. Our framework was designed to easily allow the user to rigidly align the 3D model on the image so the pose can be computed. In practice, we noticed that a rough pose estimation suffices, nevertheless, since the reprojection error will never equal zero, we slightly relax the projective constraints so that it is below 4 pixels. In fact, enforcing this constraint may lead to a alignment with a correct 2D shape projection but an aberrant 3D shape representation. In addition, the physical model ensures an out-of-plane consistency by maintaining a coherent volume altogether.

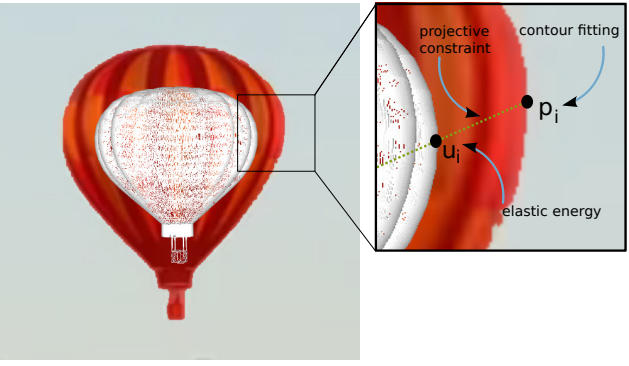

**Figure 5:** *Model-to-image alignment: energy minimization. We express the problem as an energy minimization problem where the solution is the equilibrium between the internal energy of the physical model* **u** *and the external energy corresponding to object contour fitting* **p**. *The projection constraints are included such that each 3D points $u_i$ lies on a sightline passing through camera position and 2D contour point $p_i$.*

## 6 Measuring object and scene appearance

Producing a realistic sequence $Q^*$ depends on the knowledge of the object material properties and scene illumination parameters. The literature on this subject is large [Debevec 2008] and various methods exist that can be sorted depending on the input data and scene conditions. We rely on an iterative rendering and optimization approach to recover object albedo (diffuse reflectance) and lighting parameters from the input image $I$ and the aligned model $S^*$. This enables creating a realistic rendering without visual break w.r.t the input video sequence $Q$, minimizing $\|Q - \mathscr{A}(S^*, Q_p)\|$, with $Q_p^t$ being the inpainted sequence.

Without loss of generality, object materials properties can be captured using the bi-directional reflectance function (BRDF).

$$R(s_i^*, \theta_0, \omega_0) = \int_\Omega T(s_i^*) L_i(\theta_i, \omega_i) \rho(\theta_i, \omega_i, \theta_0, \omega_0) \cos(\theta_i) d\omega_i \quad (10)$$

where $\rho$ is the BRDF of the surface, $L_i$ is the incoming radiance from direction $\omega_i$, $\theta_i$ is the angle between incoming direction and surface normal, and $T(s_i^*)$ is the texture at surface vertex $s_i^*$ of the correct geometry $S^*$. This expression is ill-posed, since we wish to estimate the BRDF from a single image of the surface, where each pixel depends on the BRDF (a four variables function), and an unknown illumination (a two variable function). In addition to a known geometry $S^*$ and texture $T$ of our object, and with the assumption of a unique distant illumination field and a Lambertian surface, the problem involves recovering the properties of $\rho$ and $L$ by minimizing the quantity

$$\left\| I_{[\text{ROI}]}^t - \sum_{i=1}^d R(s_i^*, \theta_0, \omega_0) \right\| \quad \text{with } t = t_a \quad (11)$$

where $I$ is the choosen image (corresponding to the frame $t_a$) from the sequence $Q$ and $d$ the number of pixels of the region of interest (ROI) on image $I^t$ corresponding to the projection of $S^*$ (using the whole image leads to compensation effects). The residual error is the sum of squared differences of the rendered and actual image. For the BRDF materials, we fit a Torrance-Sparrow model with 3 parameters for diffuse reflectance, 1 for specular reflectance and 1 for roughness. In addition to 3 illumination parameters for light direction.

The optimization scheme is a multi-pass process since finding all parameters at once can hardly converge. We opt for a four-pass scheme where we first find illumination parameters, then diffuse reflectance, specular reflectance, and roughness. The final pass includes all parameters using the output of each pass as initial values.

## 7 Scene Dynamics

When dealing with videos, one major challenge is to be able to produce a physically realistic motion that is synchronized with ambient dynamics. To this end we exploit videos to estimate scene dynamics that influence object's motion and we consider two parameters, the velocity **v** and the instant of collision $t_c$. These quantities are extracted by analyzing the image flow on a user-defined region of interest. We thus obtain, from a small set of successive frames, the value of **v** that we integrate in the equation of motion of the deformable object following Newton's second law:

$$\mathbf{M} \cdot \dot{\mathbf{v}} = g(\mathbf{u}, \mathbf{v}) + \mathscr{F} \quad (12)$$

where **M** is the mass matrix of the object, **v** represents the velocities and $\dot{\mathbf{v}}$ the accelerations of the element nodes, $g(\mathbf{u}, \mathbf{v})$ sums up forces that are related to the position or velocities of nodes where the function $\mathscr{F}$ gathers external force-fields (such as gravity). The moment of collision $t_c$ is detected as a sudden change of flow direction and is used as trigger to update the acceleration $\dot{\mathbf{v}}$ or any external forces $\mathscr{F}$.

Equation 12 is often solved using time-stepping techniques where time is discretized in a sequence of fixed time-steps [Anitescu et al. 1999]. Assuming a step $h = t_f - t_i$ where $t_i$ is the time at the beginning and $t_f$ at end of the step, and following an implicit (or backward) Euler integration scheme (often used as it provides increased stability compared to explicit methods especially when dealing with large time-steps), integrating equation 12 leads to:

$$\left(\mathbf{M} - h\frac{\partial g}{\partial \mathbf{v}} - h^2 \frac{\partial g}{\partial \mathbf{u}}\right) \cdot d\mathbf{v} = h^2 \frac{\partial g}{\partial \mathbf{u}} \mathbf{v}_i - h(\mathbf{g}_i + \mathbf{f}_f) \quad (13)$$

where $\mathbf{g}_i$ and $\mathbf{f}_i$ are $g(\mathbf{u}, \mathbf{v})$ and $\mathscr{F}(t)$ at time $t_i$. This enables a linear matrix-system $\mathbf{Ax} = \mathbf{b}$ where **x** is $d\mathbf{v}$. Solving this equation over time allows computing the motion of the deformable object. We choose to use an iterative algorithm to efficiently solve this problem, for instance the Conjugate Gradient iterative solver [Baraff and Witkin 1998]. This iterative method can be tuned to achieve accuracy as well as speed by controlling the number of iterations and residual error threshold.

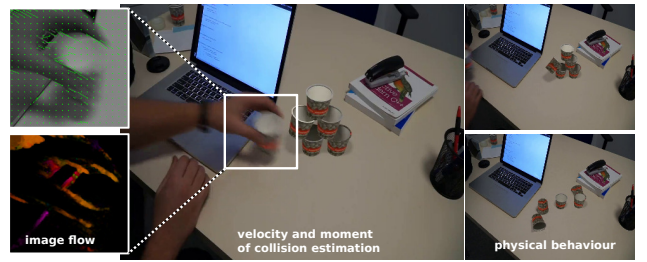

**Figure 6:** *Scene Dynamics: we estimate velocity* **v** *and instant of collision $t_c$ from the input scene by analyzing the image flow on a user-defined region of interest. Integrating these quantities to Newton's second law of motion produces a physically realistic motion that is synchronized with ambient dynamics.*

## 8 Boundary Constraints

Frequently, objects are attached to their surroundings. Physically speaking, these attachments are considered as fixed boundary conditions and are modeled by enforcing some nodes to have a null displacement. To this end, the user simply defines on the meshes the fixed regions, which lead to the binary label vector $\mathbf{q} \in \{0,1\}^n$, where $\mathbf{q}(j) = 1$ means the $j^{th}$ node is fixed and $\mathbf{q}(j) = 0$ means the $j^{th}$ node can have a free displacement. Adding constraints in equation 12 is usually performed by adding a term $\mathbf{H}^T \lambda$ where $\mathbf{H}$ is a matrix containing the constraint directions (how the vertices are constrained) and $\lambda$ is a vector of Lagrange multipliers containing the constraint force intensities and is an unknown. The matrix $\mathbf{H}$ is a sparse diagonal matrix where only fixed nodes have non-null coefficients. The equation 13 will therby be expressed following: $\mathbf{Ax} = \mathbf{b} + h\mathbf{H}(\mathbf{u})^T \lambda$ with $h$ being the integration time-step.

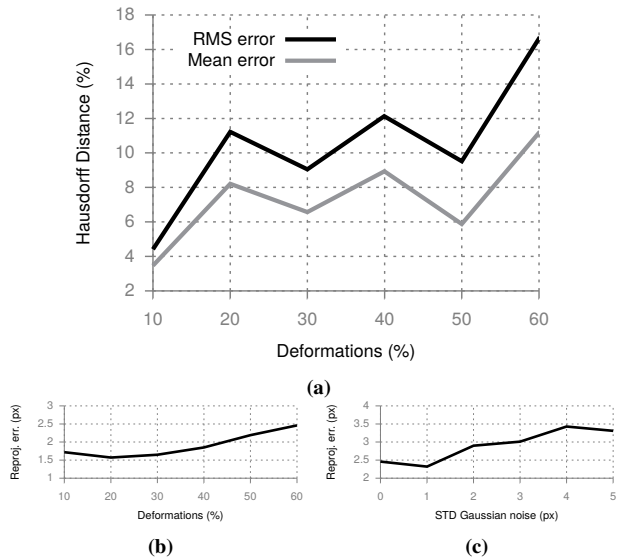

**Figure 7:** *Quantitative evaluation of the alignment and model refinement. The Hausdorff distance errors of the out-of-plane refinement are below 10% (mean) and below 12% (RMS). The error however slightly increases when deformations reach 60%. The reprojection errors are below 3 px despite the variation of the amount of deformation and the noise perturbation.*

## 9 Results

We present in this section the results obtained on different videos and images, and the experiments conducted to assess our method. We expose an evaluation of the model-to-image alignment on computer-generated data with ground and real data using visual assessment. We compute the reprojection error while varying the amount of deformations between the input 3D CAD model and the (refined) projected one. We then show various user object manipulations and physical properties editing such as pulling, copy-pasting, mass update and cutting, while taking advantage of scene dynamics, if any, to produce physically coherent motion and deformation. We finally present the user study conducted to evaluate our method, where participants were asked to rate both the ease of the alignment process and the realism of the final composition in term of appearance.

We use the framework SOFA [Faure et al. 2012] for physics-based modeling, simulation and ray-tracing techniques for rendering. We rely on the GrabCut algorithm [Rother et al. 2004] for the contour extraction and PatchMatch algorithm for image inpainting [Barnes et al. 2009].

Our dataset includes 8 videos in various environments and with different scenarios (see Figure 10), namely: *air-balloon*, *car-bridge*, *bananas*, *cup*, *eiffel-tower*, *clothsline*, *airplane-wing* and *tower-bridge*. In addition, we also show image edits on the *persistence-of-memory* painting by Salvador Dali. All examples run interactively at the same frame rate as the input video.

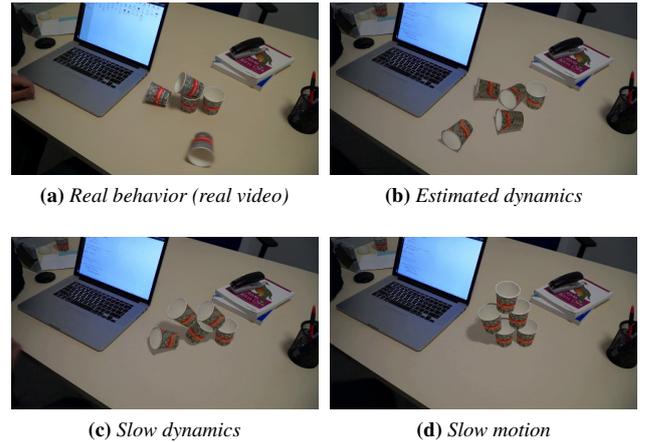

(a) *Real behavior (real video)*    (b) *Estimated dynamics*

(c) *Slow dynamics*    (d) *Slow motion*

**Figure 8:** *Scene dynamics: with a real behavior as comparison in (a), the user can choose to produce a natural physical behavior in (b) by estimating scene dynamics from image flow velocity, or to output a slower motion in (c) or a slow motion videos in (d) by tuning the velocity. All these images are taken at the same instant ($t = tc + 1$). Please see the additional material.*

### 9.1 Alignment and Model Refinement

In order to evaluate our model-to-image alignment and refinement algorithm, we used computer-generated data as ground truth where we vary the amount of deformation. That is, from an input mesh $S$, we produce several meshes $S(d)$ with different levels of deformation, where the parameter d means that the newly generated mesh is deformed by d%. The amount of deformation is measured as the average percent difference between mesh triangles' area, with $d \in \{0\%, 10\%, 20\%, 30\%, 40\%, 50\%, 60\%\}$ with $S = S^*(0)$.

The aim of the algorithm is to find $S^*(d)$ using the input mesh $S$ and the 2D projection of $S(d)$ w.r.t the camera position $\mathbf{P}$. We choose the reprojection error $\|\mathbf{P}s_i^*(d) - \mathbf{P}s_i(d)\|$ to measure in-plane alignment and the 3D Hausdorff distance (vertex-to-vertex distance) to measure out-of-plane model refinement. We also add Gaussian noise with standard deviation $n_{std} \in \{1px, 2px, 3px, 4px, 5px\}$ to the silhouette of the deformed mesh $S^*(60)$ to simulate a scattered contour that we usually obtain from a user-driven contour definition. The results are reported in Figure 7.

Our method produces low reprojection errors, i.e. below 3 px, indicating a good 3D/2D alignment in 2D space. The amplitude of deformation does not highly influence this result, where the variation between $S^*(10)$ and $S^*(60)$ is approximately 1 px. Our method also works well perturbed with noise, where the impact of these perturbations on the reprojection mean error is around 1 px with the most deformed case. The out-of-plane refinement is also relatively low with 3D mean errors below 10% and 3D RMS error below 12%.

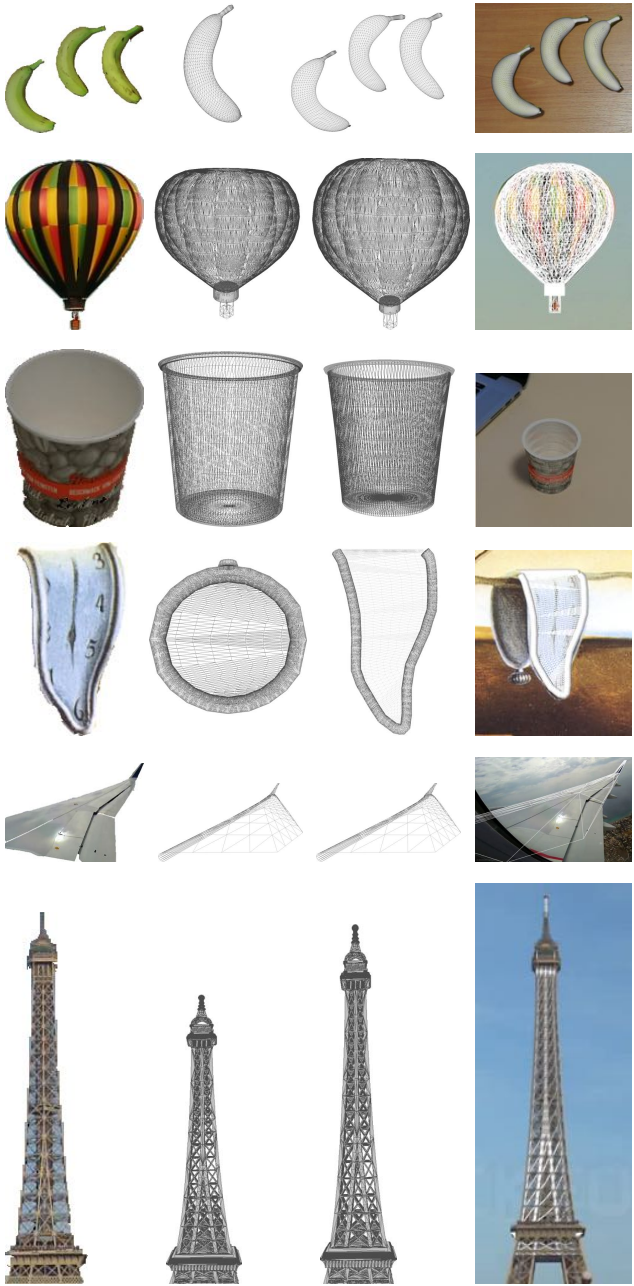

**Figure 9:** *3D Model refinement and alignment: The first column represents the user-defined contour, the second the input 3D CAD model from proxy, the third the refined model and the fourth column represents the 3D/2D alignment. (The eiffel-tower overlay shows a decimated mesh for better visualization).*

The error however slightly increases when the deformations reach 60%.

We finally use our method on real data, with results illustrated in Figure 9.

## 9.2 Object Manipulations and Physics-based Editing

**Rigid transform and copy-paste** the user can naturally translate, rotate or scale an object as well as duplicate it. In Figure 10, the *air-balloon*, the *bananas* and the *cup* are duplicated and moved all over the scene. This copy-paste implicitly implies the physical model, however, volume resolution and physical attributes can be set as desired for each object.

**Non-rigid manipulation** we enable object manipulation in a 3D manner. Most of the time these manipulations include stretching, torsion, and compression. Apart from the *cup* example, all examples in Figure 10 show non-rigid user interactions. The produced deformations highly depend on the values of $E$ the Young's modulus, and $\mu$ the Poisson ratio. For users not accustomed to using physical engines, pre-defined parameters are set according to object size and units. We also consider topological changes such as cutting or fracturing, illustrated in Figure 10 ( *bananas* and *car-bridge*).

**Collision** collisions are detected by computing a proximity distance between two or more objects in the scene. As the contact test can be computationally expensive, we only consider object vertices as primitives for the detection. Our method enables rigid/rigid collisions (*cup*), rigid/deformable collisions and deformable/deformable collisions (*bananas*). The contact response is computed following the method described here [Saupin et al. 2008]. From a user point of view, friction and minimal proximity values are parameters to be set (for instance 0.1*N* and 5*mm* respectively) as well as including/excluding colliding objects in the scene.

**Scene dynamics** recovering scene motion parameters consists on estimating velocity and moment of collisions to synchronize with scene dynamics. In the results shown, we used this principle in three examples: *air-balloon*, the *cup* and *clothsline*. In the *air-balloon* example, an initial velocity is estimated from the original object and integrated to the simulation. Assuming a linear motion, this leads to continuously translating the object with a constant speed. With the *cup* example, we estimate velocity (and thereby acceleration) to reflect object falling in a consistent manner (see Figure 8). In addition to the velocity, sudden change of motion permits obtaining instant of collision. The nature of the fall can be tuned by the user through velocity and acceleration. The time step parameter $h$ can also affect the behavior, where a small time-step will lead to transient effect (vibrations), contrary to a large time-step that will produce a smoother motion. Similarly, wind speed can be estimated from videos. In *clothsline* example, the velocity is transformed to pressure and act as external force on mesh surface. The wind speed (now represented as force in $\mathscr{F}$) can then be increased as desired by the user.

**Physical properties editing** physical attributes, whether intrinsic (related to the object, like stiffness, mass, and damping), or extrinsic (related to the scene, like gravity, or external forces), can be updated online – directly in the video. In the *air-balloon* example, we simulate deflation by reducing the stiffness from $E$ to $E \times 10^{-3}$, with $E = 2500Pa$ and $\nu = 0.45$. The deflation is undone by resetting the initial value of $E$, producing an immediate inflation. We also increase the gravity from g to g $\times$ 4, with g $= 9.8 m/s^2$ so that all objects in the scene fall. This results is similar to increasing the mass of an object as in the *car-bridge* example where the mass is increased from m to m $\times 10^3$.

### 9.3 User Study

We conducted a user study where we asked participants to evaluate both the realism of the final composition in term of appearance and the practicality of the 3D/2D alignment.

We prepared for the appearance study 10 pairs of images composed of the first frame of the original video and the corresponding output (with no deformations). The images are presented in a randomly permuted order, on a web-based form and participants are asked to classify, in less than 5 seconds, each image as *realistic* or *not realistic*. The user is forced to choose one of the two choices. The study comprised 22 participants, mostly graduated from computer science and with some computer graphics knowledge. On the 22 participants, an average of 40.33 % users classified the output image as *not realistic* where 71.08 % classified the original image as *realistic*.

For the alignment study, users are asked to rate and evaluate their ability to align the 3D model on the 2D contour by rating the practicality of the method from 1 (*not practical at all*) to 5 (*very practical*). We choose the *bananas* example (cf Figure 9, first row) where users are asked to successively perform the alignment on each contour. The study comprised 7 participants, mostly with computer graphics background that rated our method with an average of 4.35. We also compare our method with a control-point based alignment approach where we measure the time taken to align the 3D model on all projections. The contour extraction is included in the time measure for our method, where it's excluded for the control-point based approach. Using our method, the average time needed to perform the alignment is 56 seconds (*min* = 42 seconds; *max* = 91 seconds) with a maximum of of 34 seconds per projection, while the average measured time for the control-point based approach is $8:53$ minutes (*min* $= 5:05$ minutes; *max* $= 13:54$ minutes) with a maximum of 6 : 41 minutes per projection .

## 10 Limitations

The main limitations of our system reside in the adequate estimation of a 3D alignment solely from a 2D contour. This limitation represents a fundamental problem, since its under-constrained nature makes it mathematically ill-posed. Ideally, we would like to have a 3D abstraction of the scene making the alignment fully automatic and more accurate. Our work can highly benefit from recent works based on neural networks [Choy et al. 2016], [Xie et al. 2016], where 3D reconstructions are computed using learned models that extract shape priors from large collection of 3D data.

With dynamic backgrounds in videos, a temporal and coherent inpainting algorithm is necessary to produce a final composition without visual break. Although we consider this issue as a general limitation that we do not solve in our main focus, using physics-based models and scene dynamics can bring useful information to the establishment of a temporal inpainting, where object position can be predicted in time, and their silhouette can be used to compensate the new revealed parts of the video.

Interacting with objects in videos can produces occlusions, especially with newly inserted objects which can lead to visual failures. With a 3D abstract of the scene one can avoid these artifacts. However, it remains a challenging problem when dealing with a monocular camera where scene depth is unavailable.

We provide user with pre-defined initial physical parameters, computed from mesh magnitude and units. As the same way as scene dynamics, and following recent studies [Davis et al. 2016], we can exploit vibrations to estimate object's mass and stiffness to produce realistic physical output when dealing with deformable objects.

## 11 Conclusion

We presented Calipso, an interactive system for image and video editing in which the user can make physics-based 3D edits in an intuitive way. Stock 3D objects can be easily adapted to the image or video using an efficient elastic 3D-2D registration that fits the geometry to its silhouette in the image. Then, changing the object geometry, appearance and physical properties is made easy using the underlying physics-based model. At this stage of editing, the user can interact with the virtual object in many ways, from affine transformations to elastic deformations, while accounting for topological changes and collisions with the environment. In addition, it is possible to exploit time-dependent information from videos to estimate, for instance, the objects velocity and more generally to synchronize the simulation with the scene's ambient dynamics. This leads to a natural behavior of the inserted object across multiple frames of a video. We tested our approach on several examples, showing that physics-based manipulations produce visually realistic videos with both natural and unnatural physical behavior, as desired by the user. We also showed that our approach leads to low in-plane and out-of-plane alignment errors, while ensuring a significant gain in user interactions and manipulation time. We believe that a natural extension of this work would be to handle videos where the camera is in motion. This would allow to fully exploit the 3D nature of the proxy and the fact that it can be seen from different view points.

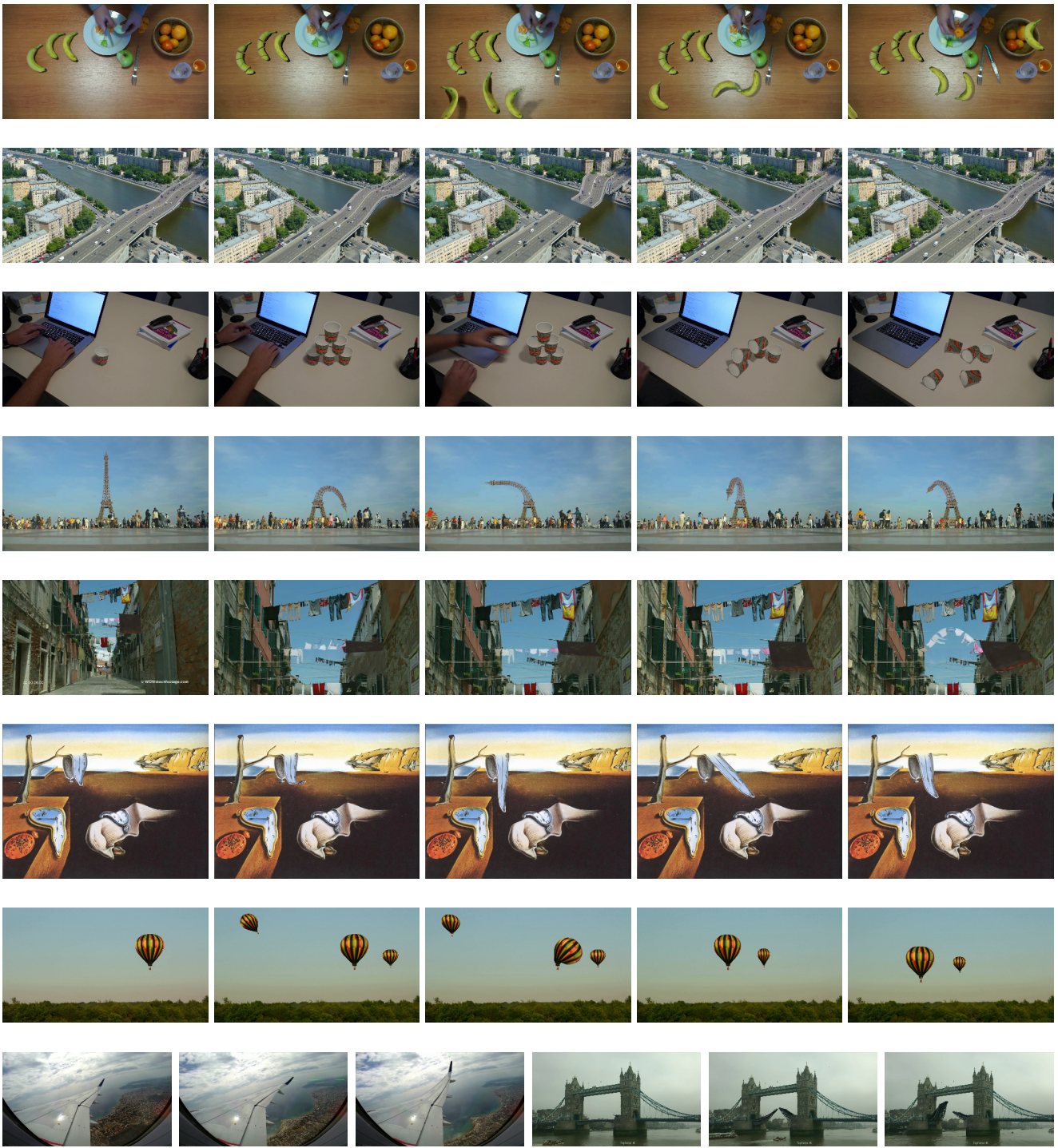

**Figure 10:** *3D physics-based manipulations of various objects in images and videos. Our method enables automatic alignment and refinement of an input 3D mesh while matching the visual characteristics of the original scene, as shown in the first image of each sequence. The user can interact with objects through a rich range of manipulations, including rigid transforms, geometry and topological changes, elastic deformations, collision and physical parameter updates such as mass, stiffness, damping, or gravity, and thereby cut the bananas and let them collide with a table, increase the mass of a car bridge and fold it up, create a pyramid of cups and make it fall, bend the Eiffel Tower, make the wind blow to wave the cloths, melt a watch in a painting, duplicate the air balloon and deform it and deflate it, stretch an airplane wing, and create an elastic Tower Bridge. [video sequences are shown in the supplemental video]*